\documentstyle[12pt,prd,aps,epsfig,preprint]{revtex}
\begin{document}
\draft
\def\ds{\displaystyle}
\date{\today}
\title{Planck Distribution in Noncommutative Space}

\author{Cem Yuce~\thanks{e-mail: e117150@metu.edu.tr}}
\address{ {\it Department of Physics , Abant Izzet Baysal University, Bolu, Turkey}} \maketitle

\begin{abstract}
In this study, we derive the Planck distribution function in
noncommutative space. It is found that it is modified by a small
quantity. It is shown that it is reduced to the usual Planck
distribution function in the commutative limit .
\end{abstract}

\thispagestyle{empty} ~~~~\\
\newpage
\setcounter{page}{1}
\section{Introduction}

In quantum mechanics, the phase space is defined by replacing the
canonical position and momentum variables with the Hermitian
operators which obey the well-known Heisenberg commutation
relations $[\hat{x}_i,\hat{p}_j]=i\hbar \delta_{i j}$. Later,
inspired by the quantum mechanics, it was suggested that one could
use the idea of space-time noncommutativity at very small length
scales to introduce an effective ultraviolet cutoff \cite{synder}
\begin{eqnarray}
[x^\mu,x^\nu]=i\Theta^{\mu \nu},
\end{eqnarray}
where $\Theta^{\mu \nu}$ is an antisymmetric tensor describing the
strength of the noncommutative effects and plays an analogous role
to $\hbar$ in usual quantum mechanics. Note that the matrix
$\Theta^{\mu \nu}$ is not a tensor since its elements are
identical in all reference frames.\\
Noncommutative field theories are constructed from the standard
field theories just by replacing the usual multiplication with the
Moyal $\star$-product in the Lagrangian density. The
$\star$-product of the two fields up to first order is defined by
\begin{equation}{\label {expansion}}
\big(\Psi \star \Phi \big)(x) = \Psi (x) \Phi (x)+\frac{i}{2}
\Theta^{\mu \nu}\partial_{\mu} \Psi (x) \partial_{\nu} \Phi (x)~.
\end{equation}
Now, let us write Hamiltonian of the particle interacting with the
radiation field in noncommutative space. The interaction
Hamiltonian in noncommutative space is assumed to be obtainable
from the standard prescription if we replace the ordinary product
with the $\star$-product. The idea is to formulate the theories in
noncommutative space as theories in commutative space and to
express the noncommutativity by an appropriate $\star$-product.
Then, the interaction Hamiltonian in noncommutative space is given
by
\begin{equation} {\label {Hcontr}}
\hat{H}_{int.}=-\frac{e}{m c} \hat{\textbf{A}} \star \textbf{p}~.
\end{equation}
Note also that $\ds{\hat{\textbf{A}}}$ in the above Hamiltonian is
the noncommutative gauge field operator. Here, caret indicates the
noncommutative quantities.\\
Gauge theory was formulated in noncommutative space
\cite{gauge1,gauge2,gauge3}. The Seiberg-Witten map allows us to
write the noncommutative fields in terms of the ordinary fields.
The Seiberg-Witten map for the gauge field is known to lowest
order in $\ds{\Theta^{\mu \nu}}$ \cite{seiberg}
\begin{equation} {\label {noncA}}
\hat{A_{\nu}}=A_{\nu}-\frac{1}{2}\Theta^{\alpha \beta} A_{\alpha}
(\partial_{\beta}A_\nu+F_{\beta \nu})~.
\end{equation}
Substitution of the equation (\ref{noncA}) into the Hamiltonian
(\ref{Hcontr}) and applying the definition (\ref{expansion})
yields
\begin{equation} {\label {Hcontr2}}
\hat{H}_{int.}=-\frac{e}{m c} \textbf{A} . \textbf{p}-\frac{ie}{2m
c}~ \Theta^{\mu \nu}
\partial_{\mu} \textbf{A}.
\partial_{\nu} \textbf{p}+\frac{e}{2m c} \Theta^{\mu \nu} \sum_{j=1}^3
A_{\mu}(\partial_{\nu}A_{j}+F_{\nu j}) p_{j}~,
\end{equation}
where $p_j$ is the $j$-th component of the momentum operator and
$A_{j}=(0,-\textbf{A})$. The first term in the above Hamiltonian
is the standard one and the other two terms
are due to the noncommutative nature of the space.\\
A few more words about the contribution to the interaction
Hamiltonian are in order. Since $\textbf{A}$ contains only the
operators $a^{\dagger}$ and $a$ for photons linearly, the second
term in Eq. (\ref{Hcontr2}) induces transitions for which one
photon is either produced or annihilated. On the other side, the
last term in the equation (\ref{Hcontr2}) induces transitions
where two photons are involved. Here, we are interested only in
one photon processes, so we restrict our consideration to the
interaction Hamiltonian which includes $\textbf{A}$ linearly. So,
we omit the last term in Eq. (\ref{Hcontr2}).\\
If we use the plane wave expansion of $\textbf{A}$ with the box
normalization condition, the second term in the above Hamiltonian
can be rewritten as
\begin{equation}
\frac{ie}{2m c}~ \Theta^{\mu \nu}
\partial_{\mu} \textbf{A}.
\partial_{\nu} \textbf{p}=\frac{ie}{2mc \hbar} \Theta^{\mu \nu}  \sum_{k
\rho} N_k~ k_{\mu} (a_{k \rho} e^{i(k.x-wt)}-a^\dagger_{k \rho}
e^{-i(k.x-wt)}) ~ \hat{\epsilon}_{k \rho} ~. \textbf{p}~p_{\nu}~,
\end{equation}
where $\ds{\hat{\epsilon}_{k \rho}}$ are the polarization vectors
($\rho=1,2$) and $\ds{N_k=\sqrt{\frac{2\pi \hbar c^2}{L^3
\omega_k}}}$. Here, we used the following identity:
$\ds{\partial_{\nu} \textbf{p}=-\frac{i}{\hbar} p_{\nu}
\textbf{p}}$.\\
Having obtained the contribution to the interaction Hamiltonian,
now we can calculate the contribution of this noncommutative term
to the first order transition matrix element.
\begin{equation}
<f,n_{k\rho}\pm 1 \mid  \frac{ie}{2m c \hbar} \Theta^{\mu \nu}
\sum_{k \rho} N_{k} k_{\mu} (a_{k \rho} e^{i (k.x-\omega
t)}-a^\dagger_{k \rho} e^{-i (k.x-\omega t)}) ~ \hat{\epsilon}_{k
\rho}~.\textbf{p}~p_{\nu}~|i,n_{k\rho}>
\end{equation}
An atom is in the initial state $|i>$ and decays into the final
state $|f>$ by emitting a photon with the wave vector $\textbf{k}$
and polarization $\rho$. It can also change it's state by
absorbing a photon with energy $\hbar \omega$. The initial and the
final of the total atom+radiation field are denoted by
$\ds{|i,n_{k\rho}>}$ and $\ds{<f,n_{k\rho}\pm 1|}$, respectively.
The number of photons is increased or decreased by one unit. Note
that there is a minus sign in front of the operator
$\ds{a^{\dagger}}$. This is because of the existence of the
derivative of the gauge field operator in Hamiltonian
(\ref{Hcontr2}). This leads to some differences between the two
theories. For example, the transition matrix elements in electric
dipole approximation are the same both for absorbtion and emission
processes in usual quantum mechanics. However, they are not equal
in noncommutative theories.\\
Let us calculate the first order transition matrix element for
emission of a photon.
\begin{equation}{\label {cem1}}
\frac{e}{mc }\sqrt{n_{k \rho}+1}~ N_k <f\mid e^{-ik.x} ~
\hat{\epsilon}_{k^{\prime}
\rho^{\prime}}~.\textbf{p}~(1+\frac{i}{2\hbar}\Theta^{\mu \nu}
k_{\mu}p_{\nu})~ \mid i> ~.
\end{equation}
Similarly, the transition matrix element for the absorbtion of a
photon is given by
\begin{equation}{\label {cem2}}
\frac{e}{mc }\sqrt{n_{k \rho}}~ N_k <f\mid e^{ik.x} ~
\hat{\epsilon}_{k^{\prime}
\rho^{\prime}}~.\textbf{p}~(1-\frac{i}{2\hbar}\Theta^{\mu \nu}
k_{\mu}p_{\nu})~ \mid i> ~.
\end{equation}

\section{Noncommutative Planck Distribution}

In this section, we derive Planck's radiation formula. It was
first derived by Einstein in 1917. We consider a sample of atoms
in thermal equilibrium. The number of atoms in state $|f>$ is
denoted by $N_f$ and for those in state $|i>$ the number $N_i$.
Transitions occur between the two states;
photons absorbed or emitted from the radiation field.\\
Equilibrium requires that time derivation of the number of atoms
is zero.
\begin{equation}
\dot{N_f}=\dot{N_i}=0~,
\end{equation}
The number of particles with energy $E$, in thermal equilibrium at
temperature $T$, is proportional to the Boltzmann factor, so
\begin{equation}{\label {papi}}
\frac{N_f}{N_i}=\exp({-\frac{E_f-E_i}{k_B T}})~,
\end{equation}
where $T$ represents temperature and $k_B$ is Boltzmann's
constant.\\
We know from quantum statistical mechanics that this ratio is also
equal to \cite{greiner}
\begin{equation}{\label {hadi}}
\frac{N_f}{N_i}=\frac{~\left( Trans.~ prob.~/~time
\right)_{emis.}}{\left( Trans.~ prob.~/~time\right)_{abs.}}~.
\end{equation}
Now, we are in a position to calculate this ratio by taking the
absolute square of the transition matrix elements
Eqs.(\ref{cem1},\ref{cem2}).
\begin{eqnarray}{\label {mamik}}
&&\frac{~\left( Trans.~ prob.~/~time \right)_{emis.}}{\left(
Trans.~
prob.~/~time\right)_{abs.}} = \nonumber\\
&&\frac{n_{k \rho}+1}{n_{k \rho}}~\times~ \frac{|<f\mid e^{-ik.x}
~ \hat{\epsilon}_{k \rho}~.\textbf{p}~ \mid
i>+~(i/2\hbar)~\Theta^{\mu \nu} k_{\mu}~<f\mid e^{-ik.x} ~
\hat{\epsilon}_{k \rho}~.\textbf{p}~p_{\nu} \mid i>|^2}{|<f\mid
e^{ik.x} ~ \hat{\epsilon}_{k \rho}~.\textbf{p}~ \mid
i>-~(i/2\hbar)~\Theta^{\mu \nu} k_{\mu}~<f\mid e^{ik.x} ~
\hat{\epsilon}_{k \rho}~.\textbf{p}~p_{\nu} \mid i>|^2} ~.
\end{eqnarray}
Let us compute the term with $\ds{\Theta^{\mu \nu}}$ further by
using the electric dipole approximation method $(e^{\pm ikx}
\approx 1)$.
\begin{eqnarray}{\label {hahi}}
\Theta^{\mu \nu} k_{\mu} <f\mid  ~ \hat{\epsilon}_{k
\rho}~.\textbf{p}~p_{\nu}~ \mid i> ~=~\Theta^{\mu \nu} k_{\mu}~
\sum_{l}~ <f\mid \hat{\epsilon}_{k \rho}~.\textbf{p} \mid l>
<l\mid p_{\nu} \mid i> ~,
\end{eqnarray}
where $\ds{p_{\nu}=(p_0,-\textbf{p})}$.\\
If $\nu=0$, then $\ds{<l\mid p_{0}\mid i> =\frac{E_i}{c}
\delta_{li}}$. If $\nu \neq 0$, then $\ds{<l\mid p_{j}\mid i> =i
m\omega_{li} <l|x_{j}|i>}$, where $j=1,2,3$. Then, the equation
(\ref{hahi}) can be rewritten
\begin{eqnarray}
\frac{i m\omega_{fi} E_i}{c}~ \Theta^{\mu 0} k_{\mu} ~ <f\mid
\hat{\epsilon}_{k \rho}~. \textbf{x}  \mid i>- m^2
\sum_{l}~\Theta^{\mu j} k_{\mu}~ \omega_{fl}~ \omega_{li}~ <f\mid
\hat{\epsilon}_{k \rho}~. \textbf{x} \mid l> <l\mid x_j \mid i>~.
\end{eqnarray}
The elements of $\ds{\Theta^{\mu \nu}}$ are very small
\cite{la3,small} ($\ds{\Theta^{\mu \nu}<<1}$). So, we can neglect
the square terms of $\Theta^{\mu\nu}$. Let us define $\epsilon$
which includes $\ds{\Theta^{\mu \nu}}$ linearly as
\begin{equation}
\epsilon= \frac{~\frac{m}{2\hbar}~\Theta^{\mu j} k_{\mu} \sum_{l}~
\omega_{fl} \omega_{li}~ <f\mid \hat{\epsilon}_{k \rho}.
\textbf{x} \mid l> <l\mid x_j \mid i>}{\omega_{fi} <f\mid
\hat{\epsilon}_{k \rho}~. \textbf{x} \mid i>}~.
\end{equation}
Then the ratio Eq. (\ref{mamik}) is approximated as
\begin{equation}{\label {kaki}}
\frac{~\left( Trans.~ prob.~/~time \right)_{emis.}}{\left( Trans.~
prob.~/~time\right)_{abs.}} \simeq \frac{n_{k \rho}+1}{n_{k
\rho}}~\times~ \frac{1-2\epsilon}{1+2\epsilon}\simeq \frac{n_{k
\rho}+1}{n_{k \rho}}~\times~(1-4\epsilon) ~,
\end{equation}
At the last step, we used the following relation
$\ds{(\frac{1}{1+\epsilon}\simeq 1-\epsilon)}$. Substitution of
the equations (\ref{papi},\ref{kaki}) into the definition
(\ref{hadi}) leads to the Planck distribution function in
noncommutative space.
\begin{equation}{\label {cemyuce}}
n_{k \rho}=\frac{1}{\kappa \exp({\frac{\hbar \omega_k}{k_B
T}})-1}~,
\end{equation}
where $\kappa$ is given by
\begin{equation}
\kappa=\frac{1}{1-4\epsilon}\simeq 1+4\epsilon~.
\end{equation}
In commutative limit ($\ds{\Theta^{\mu \nu}\rightarrow 0}$,
$\kappa\rightarrow 1$), the Eq. (\ref{cemyuce}) is reduced to the
usual Planck distribution function.\\
Cosmic black body radiation may be one of the experimental test of
noncommutative theories. The remnants of the intense radiation
field produced in the beginning can be presented as a black body
radiation. The value of the physical parameters like energy and
temperature should be shifted because of the parameter $\kappa$ in
Eq. (\ref{cemyuce}), if noncommutativity of the space plays a role
in the first times of big bang. One can study many new interesting
physical phenomena with this noncommutative Planck distribution
function Eq. (18).

Up to now, we have not said anything about the elements of the
noncommutative parameter $\Theta^{\mu \nu}$. Actually, there is no
agreement in the literature on how these elements are constructed
explicitly. While many researchers set $\ds{\Theta^{0i}=0}$ to
avoid problems with unitarity and casuality \cite{R8}, the others
assume that the components of $\ds{\Theta^{\mu \nu}}$ are constant
over cosmological scales, in any given frame of reference there is
a special noncommutative direction \cite{la2}.
Hewett-Petriello-Rizzo \cite{R15} parametrized them with the three
different angles. In \cite{ek2}, $\ds{\Theta^{\mu \nu}}$ is
interpreted as a background B-field.

\end{document}